\documentclass[prb,groupedaddress,twocolumn]{revtex4}

\usepackage{amsmath}
\usepackage{graphicx}% Include figure files
\usepackage{dcolumn}% Align table columns on decimal point
\usepackage{bm}% bold math
\usepackage{alltt}% 

\newcommand{\mX}{\mathcal{X}}

\begin{document}

%\title{An accidental derivation of metadynamics: Error estimates and exact results}
\title{A parameter-free metadynamics}
\author{Bradley M. Dickson}
\email{bmdickso@purdue.edu}
\affiliation{Medicinal Chemistry and Molecular Pharmacology, Purdue University, 240 S. Martin Jischke Drive, West Lafayette, IN 47907-1971}
%\author{He Huang}
%\email{huang23@purdue.edu}
%\affiliation{Purdue University...}
%\author{Carol B. Post}
%\email{cbp@purdue.edu}
%\affiliation{Purdue University...}

\date{\today}

\begin{abstract}
We present a unique derivation of metadynamics. 
The starting point 
for the derivation is an on-the-fly reweighting scheme but through an approximation we recover the standard 
metadynamics and the well-tempered metadynamics in a general form while never appealing to the extended Lagrangian framework. 
This work leads to a more robust understanding of the error in the computed free energy 
than what has been obtained previously. Moreover, a formula for the exact free energy is introduced. The formula 
can be used to post-process any existing well-tempered metadynamics data allowing one, in principle, to obtain 
an exact free energy regardless the metadynamics parameters. 

\end{abstract}

\maketitle
The last decade has seen the introduction of a number of adaptive biasing techniques developed for free energy 
computation.\cite{bbp08,lrsbook,mb06,ens10} 
In practice, these techniques accumulate a biasing potential or a biasing force during trajectory evolution. The goal is to specify 
a biasing force or potential that will effectively flatten the free energy landscape and enhance sampling. 
Here we focus on a particular adaptive biasing potential 
(ABP) method called metadynamics, introduced in references \onlinecite{lp02} and \onlinecite{mlp03}. Metadynamics is an ABP method 
that has been widely used for chemical, solid state and biological systems.\cite{lg08} 

One area of metadynamics that can still be improved upon is that of the dependence of the error in the computed free energy on the metadynamics 
parameters.\cite{lrgcm05,blp06,br06} Metadynamics requires one to specify several system dependent parameters. 
The parameters are the energy rate, which 
is the product of the Gaussian height and the deposition period, and the Gaussian width.\cite{lrgcm05} 
The error associated with these is mostly well understood, with one exception: The Gaussian width. 
At present, the analytic error estimates 
suggest that, for fixed computational time, 
the error decreases as the Gaussian width increases\cite{lrgcm05,blp06} while numerical experiments reveal that 
the error will actually increase in the limit of large Gaussian widths and that the empirically derived 
error estimate will break down\cite{lrgcm05}. 

Below, we derive metadynamics in a novel way. %, albeit by accident. 
Originally, 
the goal was to propose an adaptive biasing scheme 
with 
on-the-fly reweighting (as in references \onlinecite{mb06,ens10}) 
and then, by an approximation, show that the reweighting factors could be 
removed. The problem with the on-the-fly reweighting methods is 
that they are difficult to study formally because of the time-dependent reweighting. The hope was to introduce an 
ABP method that was free from these reweighting factors, allowing it to be studied rigorously like metadynamics and adaptive biasing 
force methods.\cite{lrs08} However, this process leads one directly to metadynamics, 
both the well-tempered and standard forms. This derivation %accident 
accentuates an 
interpretation that leads to an understanding of the error incurred by choosing a finite Gaussian width and ultimately affords an 
exact expression for the free energy, independent of the Gaussian width. In other words, a formula follows from this interpretation that 
is exact even for finite Gaussian width. This formula could be used to post-process any existing well-tempered metadynamics data. 
We also stress that the biased dynamics can be cast in a form consistent with the dynamics studied in reference \onlinecite{lrs08}; 
The biasing force can be expressed as an average taken over replicas of the system rather than an average over time. It is expected 
that because of this, it should be possible to obtain convergence results for the well-tempered metadynamics without assuming 
``instantaneous equilibration'' of the dyanamics. 

Below we present the derivation by first introducing the on-the-fly reweighting strategy and then introducing our approximations. Once 
metadynamics is recovered we present an error estimate and an exact formula for free energy.

Let $x$ be a single configuration in the 
$n$-dimensional configuration space $\mX$ of some interesting dynamical system. For this system, assume $N$ collective variables (CVs) 
have been specified and further suppose that 
the CVs are good descriptors of the interesting features of $\mX$. The CV space is $\Omega$.
Let $\xi$ denote a point in $\Omega$. 
Following reference \onlinecite{ens10}, let us define the mollified 
free energy (up to an arbitrary constant $\zeta$) 
\begin{eqnarray}
  \label{molDOS}
 &\zeta e^{-\beta A_{\alpha}(\xi)}=& \\ \nonumber
&Z^{-1}\int_{\mX} \delta_{\alpha}(\xi(x)-\xi) \, e^{-\beta V(x)} \, dx, &
\end{eqnarray} where 
\begin{equation}\delta_{\alpha}(\xi)=\exp\left(-\frac{|\xi|^2}{\alpha^2}\right). 
\end{equation} 
%\begin{eqnarray}&&\delta_{\alpha}(\xi)=\exp\left(-\frac{|\xi|^2}{\alpha^2}\right)\nonumber \\
%&&\partial_{\xi_i}\delta_{\alpha}(\xi(x_t)-\xi) = \frac{2(\xi_i(x_t)-\xi_i)}{\alpha^2}\delta_{\alpha}(\xi(x_t)-\xi).
%\end{eqnarray} 
Let $Z$ absorb the normalization of $\delta_{\alpha}$. In the limit $\alpha\rightarrow 0$ this is the exact free energy. When 
$\alpha$ is finite, the exact free energy can in principal be recovered via deconvolution\cite{ens10} and we will later 
make use of this.

In practice, equation \eqref{molDOS} would be computed via trajectory averages. 
Let $x_t$ be a trajectory solving 
the following Langevin equation, 
\[d\dot{x}_t = -f\dot{x}_t dt -\nabla V(x_t)dt + \sqrt{2f\beta^{-1}}dB_t \] 
where we assume it to be ergodic with respect to the Boltzmann density on $\mX$, 
$f$ is the Langevin friction coefficient 
and $\beta$ is the inverse temperature. The random force $dB_t$ is given by the increments of 
a Brownian motion. 

Defining the population at $\xi$ 
\begin{equation}
g(\xi,t) = \displaystyle\int_0^t \delta_{\alpha}(\xi(x_s)-\xi)
\, ds
\end{equation} and 
\begin{equation}
Z_t = \displaystyle \int_{\Omega} g(\xi,t)d\xi, 
\end{equation} 
the free energy 
may be computed from a long trajectory $x_t$ as 
\begin{equation}
  \label{traj-DOS}
  \zeta \,e^{-\beta A_{\alpha}(\xi,t)}=Z_t^{-1} g(\xi,t), 
\end{equation} where at $t=0$ 
\begin{equation}\label{initi}\displaystyle 
Z_0^{-1} g(\xi,0) = \frac{\delta_{\alpha}(\xi(x_0)-\xi)}{\alpha\sqrt{\pi}}.\end{equation}

Now, we propose the following on-the-fly reweighting scheme to compute equation \eqref{traj-DOS} via an adaptive biasing potential. 
%If there are large free energy barriers in $\Omega$, equation \eqref{traj-DOS} will converge only very slowly. What we propose, 
%of course, is 
%to bias the dynamics. 
%The Langevin dynamics are to be given by equation \eqref{bLE}, where $V_b$ is an adaptive biasing potential. 
%Ideally, the bias potential is designed 
%so that in the long-time limit, the free energy landscape is at least approximately flattened. 
In this case we propose the biasing potential 
\begin{equation}
  \label{Cbias}
  e^{\beta V_b(\xi,t)} = \left(cg(\xi,t)+1\right)^b \, 
\end{equation} where $b$ and $c$ are scalars. $b$ controls the strength of the bias and $c>0$ is a coupling parameter with units of inverse time. 
With the bias defined this way, the initial conditions for the biased dynamics reduce to those of the unbiased case and $0 \leq V_b$. 

Using this bias potential, one can write the mollified free energy 
as time averages over Langevin trajectories driven by $V+V_b$ 
with on-the-fly reweighting, 
\begin{equation}
  \label{btraj-DOS}
  \zeta e^{-\beta A_{\alpha}(\xi,t)}= Z_t^{-1}g(\xi,t), 
\end{equation} where
\begin{equation}\label{GEE}
g(\xi,t) = \displaystyle\int_0^t \delta_{\alpha}(\xi(x_s)-\xi)e^{\beta V_b(\xi(x_s),s)}
\, ds
\end{equation} and 
\begin{equation}
Z_t = \displaystyle \int_{\Omega} g(\xi,t)d\xi. 
\end{equation} 
It is natural for the development of the ABP method to 
terminate here; these equations could be implemented as-is. See reference \onlinecite{ens10} for example. 

In this case one can go a bit further, however, by introducing an approximation to equation \eqref{GEE}. The idea 
behind the following manipulations is to derive an expression for $g(\xi,t)$ in which the factor $\exp[\beta V_b]$ does not appear. 
Once obtained, we recover metadynamics. 
Using a first-order expansion $V_b(\xi(x_t),t) \approx V_b(\xi,t)+V_b'(\xi,t)(\xi(x_t)-\xi)$ and an expansion of $e^x$, 
we obtain 
\begin{eqnarray}\label{GEE2}
&&g(\xi,t) %= \displaystyle\int_0^t \delta_{\alpha}(\xi(x_s)-\xi)e^{\beta V_b(\xi(x_s),s)}
%\, ds\nonumber \\
\approx\displaystyle\int_0^t \delta_{\alpha}(\xi(x_s)-\xi)e^{\beta V_b(\xi,s)}e^{\beta V_b'(\xi,s)(\xi(x_s)-\xi)}
\, ds\nonumber \\
&&\approx \displaystyle\int_0^t \delta_{\alpha}(\xi(x_s)-\xi)e^{\beta V_b(\xi,s)}( 1+ \beta V_b'(\xi,s)(\xi(x_s)-\xi))
\, ds\nonumber \\
&&=\displaystyle\int_0^t \delta_{\alpha}(\xi(x_s)-\xi)e^{\beta V_b(\xi,s)}ds + \Delta(\alpha). 
\end{eqnarray} 
The last term is roughly the error between the populations computed by equation \eqref{GEE} and the population 
computed with the first term in the last line of equation \eqref{GEE2}, 
\begin{equation}\label{error1}
\Delta(\alpha) = \frac{\alpha^2\,\beta}{2}\displaystyle \int_0^t \partial_\xi \delta_{\alpha}(\xi(x_s)-\xi)V_b'(\xi,s)e^{\beta V_b(\xi,s)}ds.
\end{equation} 

Ignoring $\Delta(\alpha)$, the final line in equation \eqref{GEE2} can be seen as an integral solution to the differential equation 
\[
\frac{dg(\xi,t)}{dt} = \delta_{\alpha}(\xi(x_t))-\xi)e^{\beta V_b(\xi,t)}, 
\] 
which can be solved via separation of variables 
\begin{equation}\label{sol1}
g(\xi,t) = \frac 1 c \, \displaystyle \left( \left( c(1-b)\int_0^t\delta_{\alpha}(\xi(x_s)-\xi)ds
+1\right)^{\frac{1}{1-b}}-1\right).  
\end{equation} 
To avoid the possibility of complex valued $g$, we restrict $c>0$ and $b\leq 1$. We can now express $g$ without evaluating 
$\exp[\beta V_b]$. 

Equations \eqref{Cbias} and \eqref{sol1} 
combine to give the bias potential 
\begin{equation}\label{meta}
V_b(\xi,t) = \beta^{-1} \frac{b}{1-b}\ln[c\,(1-b)\,\displaystyle \int_0^t \delta_{\alpha}(\xi(x_s)-\xi)ds +1]. 
\end{equation} 
In the limit $b\rightarrow 1$ the bias 
reduces to 
\begin{equation}\label{limmeta}
V_b(\xi,t) = \beta^{-1} c\,\displaystyle \int_0^t \delta_{\alpha}(\xi(x_s)-\xi)ds. 
\end{equation} 
These bias potentials are the well-tempered\cite{bbp08} and standard metadynamics\cite{lp02} bias potentials, respectively. 
Indeed, making the substitutions 
$\omega = \beta^{-1} c\,b$ and $\Delta T = \beta^{-1} b /(1-b)$ 
in equation \eqref{meta}, we find the well-tempered metadynamics, 
\begin{equation}V_b(\xi,t) = \Delta T\ln \left(\frac{\omega}{\Delta T}\,\displaystyle \int_0^t \delta_{\alpha}(\xi(x_s)-\xi)ds +1\right). 
\end{equation} 
Equation \eqref{limmeta} is the standard metadynamics with an energy rate of $\beta^{-1} \,c$. 
We have derived metadynamics via the approximations in equation \eqref{GEE2} having started with an on-the-fly reweighting scheme. 
%At this point we expect the error to be proportional to $\alpha^2$. We will validate this using a simple test case below. 

To derive an exact expression for the free energy when $0\leq b< 1$, it will be useful to work with the 
biasing force and with averages taken over independent replicas of the dynamics, rather than averages in time. 
The gradient of equation \eqref{meta} is 
\begin{equation}\label{btraj-grad}
\frac{\partial V_b(\xi,t)}{\partial \xi_i} = 
\frac{c\, b\, \beta^{-1}\,\displaystyle\int_0^t \partial_{\xi_i}\delta_{\alpha}(\xi(x_s)-\xi)\, ds}
{1+c\,(1-b)\,\displaystyle\int_0^t\delta_{\alpha}(\xi(x_s)-\xi)\, ds},  
\end{equation} where 
\begin{equation}
\partial_{\xi_i}\delta_{\alpha}(\xi(x_t)-\xi) = \frac{2(\xi_i(x_t)-\xi_i)}{\alpha^2}\delta_{\alpha}(\xi(x_t)-\xi).
\end{equation}

In the long-time limit, 
\begin{equation}\label{btraj-long}
\frac{\partial V_b(\xi,t)}{\partial \xi_i} = 
\frac{b\, \beta^{-1}\,\displaystyle\int_0^t \partial_{\xi_i}\delta_{\alpha}(\xi(x_s)-\xi)\, ds}
{\,(1-b)\,\displaystyle\int_0^t\delta_{\alpha}(\xi(x_s)-\xi)\, ds}.  
\end{equation} Notice that in the long-time limit $c$ vanishes. This implies that for well-tempered metadynamics the 
Gaussian height will not impact the long-time accuracy of the computation.

Noting the $i-$th replica of the biased dynamics with $x_t(i)$, 
the replica density can be defined 
\[ \psi(x,t) = \displaystyle \lim_{M\rightarrow \infty}\frac{1}{M} \displaystyle \sum_{i=1}^M 1_{[x_t(i) = x]}. \]
Each replica is an independent solution to the biased Langevin-equation 
\begin{equation}\label{bLE}d\dot{x}_t = -f\dot{x}_t dt -\nabla(V(x_t)+V_b(\xi(x_t)))dt + \sqrt{2f\beta^{-1}}dB_t. \end{equation}
Equation \eqref{btraj-long} can be cast as the following replica average 
\begin{equation}\label{pathint}
\frac{\partial V_b(\xi,t)}{\partial \xi} =% \frac{\beta^{-1}b \displaystyle \sum_{i=1}^M \partial_{\xi}\delta_{\alpha}(\xi(x_t(i))-\xi)}{(1-b) 
%\displaystyle \sum_{i=1}^M \delta_{\alpha}(\xi(x_t(i))-\xi)}\nonumber \\
\frac{\beta^{-1}b \displaystyle \int_{\mX} \partial_{\xi}\delta_{\alpha}(\xi(x)-\xi) \psi(x,t) dx}
{(1-b) \displaystyle \int_{\mX} \delta_{\alpha}(\xi(x)-\xi) \psi(x,t) dx}. 
\end{equation} 
%Integrating by parts one has(see lemma 7 of reference \online
%\begin{eqnarray} \label{ABFs}
%&&\frac{\partial V_b(\xi,t)}{\partial \xi} 
%=\frac{\beta^{-1}b \displaystyle \int_{\mX} \partial_{\xi}\delta_{\alpha}(\xi(x)-\xi) \psi(x,t) dx}
%{(1-b) \displaystyle \int_{\mX} \delta_{\alpha}(\xi(x)-\xi) \psi(x,t) dx} \nonumber \\
%&&= -\frac{b}{1-b}\frac{\beta^{-1} \,\displaystyle \int_{\mX} \delta_{\alpha}(\xi(x)-\xi) \mL \psi(x,t) dx}
%{ \displaystyle \int_{\mX} \delta_{\alpha}(\xi(x)-\xi) \psi(x,t) dx}, 
%\end{eqnarray} 
%where the operator $\mL$ is 
%\begin{equation}
%\mL = \frac{\nabla \xi(x) \cdot \nabla}{|\nabla \xi(x)|^2} + \text{div} \left(\frac{\nabla \xi(x)}{|\nabla \xi(x) |^2}\right). 
%\end{equation} 

Assuming that when $t\rightarrow \infty$, 
the method converges and $\psi(x,\infty) \propto \exp[-\beta (V(x)+V_b(\xi(x),\infty))]$ so that 
\begin{eqnarray}\label{mean}
%&&\frac{\partial V_b(\xi,\infty)}{\partial \xi} =\nonumber\\
&&\frac{\beta^{-1}\displaystyle \int_{\mX} \partial_{\xi}\delta_{\alpha}(\xi(x)-\xi) e^{-\beta(V(x)+V_b(\xi(x),\infty))} dx}
{\displaystyle \int_{\mX} \delta_{\alpha}(\xi(x)-\xi) e^{-\beta(V(x)+V_b(\xi(x),\infty))} dx} \nonumber \\
&&= \frac{\beta^{-1} \,\displaystyle \int_{\Omega} \partial_{\xi}\delta_{\alpha}(\bar{\xi}-\xi) e^{-\beta(A(\bar{\xi})+V_b(\bar{\xi},\infty))} d\bar{\xi}}
{ \displaystyle \int_{\Omega} \delta_{\alpha}(\bar{\xi}-\xi) e^{-\beta(A(\bar{\xi})+V_b(\bar{\xi},\infty))} d\bar{\xi}}  \nonumber \\
&&=-\frac{\partial \mu_{\alpha}(\xi,\infty)}{\partial \xi}, 
\end{eqnarray} where we have defined the mollified free energy in the biased ensemble $\mu_{\alpha}$. 
For finite $\alpha$, the biasing force is thus a rescaling of the mollified mean force in the biased ensemble and 
$V_b = -b\,\mu_{\alpha}/(1-b)$ up to an additive constant. This interpretation of the biasing force in well-tempered metadynamics 
leads directly to an exact expression for the free energy.

Consider the exact mean force in the biased ensemble, 
\begin{eqnarray}
&&-\frac{\partial \mu(\xi^*,\infty)}{\partial \xi^*} = 
\frac{\beta^{-1} \,\displaystyle \int_{\Omega} \partial_{\xi}\delta(\xi-\xi^*) e^{-\beta(A(\xi)+V_b(\xi,\infty))} d\xi}
{ \displaystyle \int_{\Omega} \delta(\xi-\xi) e^{-\beta(A(\xi)+V_b(\xi,\infty))} d\xi}\nonumber \\
&&= -(\nabla_{\xi^*} A(\xi^*)+\partial_{\xi^*}V_b(\xi^*,\infty))\nonumber \\
&&= -\nabla_{\xi^*} A(\xi^*)+\frac{b}{1-b}\partial_{\xi^*}\mu_{\alpha}(\xi^*,\infty).  
\end{eqnarray} 
The exact free energy to an unimportant constant is therefore, 
\begin{equation}\label{exact}
A(\xi) = \mu(\xi,\infty) + \frac{b}{1-b}\mu_{\alpha}(\xi,\infty). 
\end{equation} 
What makes this expression useful is that $\mu$ can be obtained by a deconvolution of $\mu_{\alpha}$ 
and the known Gaussian $\delta_{\alpha}$. 
%Denoting 
%the deconvolution with $\delta^{-1}_{\alpha}$, the exact free energy can be written in terms of the computed quantity $\mu_{\alpha}$ 
%\begin{equation}\label{decon}
%A(\xi) = -\beta^{-1}\ln\left(\delta^{-1}_{\alpha}\ast e^{-\beta\mu_{\alpha}(\xi,\infty)}\right) + \frac{b}{1-b}\mu_{\alpha}(\xi,\infty). 
%\end{equation} 

If one ignores the deconvolution and assumes $\mu \approx \mu_{\alpha}$, 
\begin{equation}\label{error}A(\xi) \approx -\frac{1}{b}V_b(\xi,\infty)\end{equation} 
where the error associated with this approximation is roughly, 
\begin{equation}\label{order}
\beta^{-1}\ln\left(1+\frac{\alpha^2}{4}\big[(\beta\mu'(\xi,\infty))^2-\beta\mu''(\xi,\infty)\big]\right). 
\end{equation} We have used $V_b = -b\,\mu_{\alpha}/(1-b)$ in equation \eqref{error}. 
This error estimate was also given in reference \onlinecite{ens10} (see equation (21) there). 
This estimate is found by making a Taylor series expansion of $\exp[-\beta \mu(\xi,\infty)]$ in 
\begin{equation}\label{error2}
e^{-\beta \mu_{\alpha}(\xi^*,\infty)} = \int_{\Omega} \delta_{\alpha}(\xi-\xi^*)e^{-\beta\mu(\xi,\infty)}d\xi
\end{equation} and keeping terms up to the second moment of $\delta_{\alpha}$. Notice that the error in well-tempered 
metadynamics is related to a convolution of the configurational density in the biased ensemble. In reference \onlinecite{ens10} the 
error was due to a convolution of the configurational density in the unbiased ensemble. 

In practice one needs the histograms 
\begin{eqnarray}
&&h(\xi,t) = \displaystyle\int_0^t\delta_{\alpha}(\xi(x_s)-\xi)\, ds\nonumber \\
&&h_b(\xi,t) = h(\xi,t)^{\frac{b}{1-b}}
\end{eqnarray} and the exact free energy can be computed with 
\begin{equation}\label{decon}
A(\xi,t) = -\beta^{-1}\ln\left(\delta^{-1}_{\alpha}\ast h(\xi,t)\times h_b(\xi,t)\right) 
\end{equation} where we use $\delta_{\alpha}^{-1}\ast h$ to indicate a deconvolution. 

%The metadynamics is 
%optimized by choosing a large Gaussian width where the primary caution is that too great of a width will result in a low resolution estimate 
%of $A(\xi)$.\cite{lrgcm05,blp06} We see here that the loss of resolution is related to a convolution of the biased configuration density on $\Omega$ 
%with $\delta_{\alpha}$. 
%Just as in reference 
%\onlinecite{ens10}, one can relate $\alpha^2$ and the local curvature of $\mu$ via equation \eqref{error}. 
%The more highly curved and feature rich $\mu$ is, the smaller 
%$\alpha$ will need to be in order to control the error introduced by ignoring the deconvolution. 

By now it has been demonstrated that metadynamics is a powerful computational tool\cite{lg08}, so we use a very simple model here to 
demonstrate that the formula given in equation \eqref{decon} can be used to compute an accurate free energy even for large $\alpha$. 
The Richardson-Lucy scheme\cite{r72,l74} was used for the deconvolution, as described in 
reference \onlinecite{ens10}. 

We apply the method with $\xi(x) = x$ and $A(x) = V(x) = x^4-x^2+0.25$. 
The dynamics are 
given by equation \eqref{bLE} 
where $f=1/2dt$, $k_BT = .25/10$ (one tenth of the 
barrier height), the particle mass 
is unity. 
The Langevin integrator from reference \onlinecite{at87} was used to evolve the trajectory. 
The coordinate $x$ is discretized from $x=-2$ to 
$x=2$ into $400$ bins. 
At $t=0$  the initial phase point is $(x,\dot{x}) = (\sqrt{0.5},0)$. 

Here we implement the above stated grid-based metadynamics for a single trajectory 
with $b=0.8$ and $c=1/dt$. The histogram $h(\xi,t)$ defined above and its derivative  
\begin{equation}h'(\xi,t) = \displaystyle\int_0^t \partial_{\xi_i}\delta_{\alpha}(\xi(x_s)-\xi)\, ds 
\end{equation} 
are computed on the grid of values $\xi$, where 
at each timestep the trajectory makes a contribution to all grid points. 
The gradient of the bias potential is given by equation \eqref{btraj-grad}. 
At the end of the simulation we apply equation \eqref{decon} to remove the impact of finite $\alpha$. 

\begin{figure}
\includegraphics[width=\columnwidth]{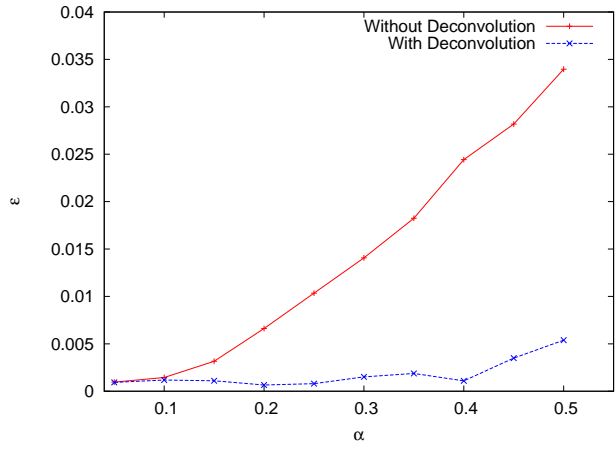}
\caption{Error as a function of $\alpha$ with and without deconvolution.
}\label{result3}
\end{figure}

We evolve for $10^7$ dynamical steps and compute the error $\epsilon =\sum_i |A(\xi_i)-\hat{A}(\xi_i,t)|d\xi/4$ where $i$ runs over all bins 
such that $A(\xi_i)<\beta^{-1}$ and $4$ is the length of the grid. 
We adopt this condition from reference \onlinecite{lrgcm05}. The error is shown in figure \ref{result3} where $\hat{A}$ is either 
the estimate of $A$ with or without deconvolution. 
The two are clearly labeled in figure \ref{result3} and we find that the deconvolution makes a significant improvement to the 
computed free energy. 
The grid spacing should satisfy $d\xi << \alpha$ so that the $\delta_{\alpha}$ are well represented on the grid. If the grid is too coarse, 
one can expect an increase in error as $\alpha$ decreases. 

In conclusion, we have presented a derivation of metadynamics that leads to an understanding of the error associated with finite $\alpha$ and 
a formula for removing this error. We have demonstrated that the formula in equation \eqref{decon} can indeed be used to correct the computed 
free energy even when $\alpha$ is large. 
Equation \eqref{decon} can be used to post porcess any existing well-tempered metadynamics data 
to remove the blurring related to using a finite $\alpha$. 
In hindsight, it appears that both equation \eqref{decon} and the error in equation \eqref{order} 
should follow straight from the presentation of metadynamics in reference \onlinecite{bbp08}. One only needs to notice that the biasing force is related 
to the mollified mean force in the biased ensemble. 

The author acknowledges Robert D. Skeel and Carol B. Post for the freedom to explore this topic and He Huang for noticing that 
given the interpretation of equation \eqref{mean}, 
equation \eqref{exact} is trivial to deduce. 
This work was supported by NIH grant number R01 GM 083605. Bevan Elliott and Paul Fleurat-Lessard are thanked for a careful reading of the manuscript.

%\begin{figure}
%\includegraphics[width=\columnwidth]{1D-2.eps}
%\caption{$V(x)$ is shown in red and two well-tempered simulations with different $c$ are shown in green and blue. }\label{result2}
%\end{figure}

\bibliography{abp.bib}
\end{document}